\documentstyle[preprint,aps]{revtex}

\input epsf         
\epsfverbosetrue    

\begin{document}

\draft

\title{Surface Contribution to Raman Scattering from Layered Superconductors}

\author{W. C. Wu and A. Griffin}
\address{Department of Physics, University of Toronto,
         Toronto, Ontario, Canada, M5S 1A7} 

\date{\today}

\maketitle

\begin{abstract} 
Generalizing recent work, the Raman scattering intensity 
from a semi-infinite superconducting superlattice is calculated
taking into account the surface contribution to the density response functions.
Our work makes use of the formalism of Jain and Allen developed for
normal superlattices. The surface contributions are shown
to strongly modify the bulk
contribution to the Raman-spectrum line shape below $2\Delta$, and also may
give rise to additional surface plasmon modes above $2\Delta$. 
The interplay
between the bulk and surface contribution is strongly dependent
on the momentum transfer $q_\parallel$ parallel to layers. However, we argue
that the scattering cross-section for the
out-of-phase phase modes (which arise from interlayer Cooper pair tunneling)
will not be affected and thus should be the only 
structure exhibited in the Raman spectrum
below $2\Delta$ for relatively large $q_\parallel\sim 0.1\Delta/v_F$. The
intensity is small but perhaps observable.
\end{abstract}

\vskip 0.1 true in

\pacs{PACS numbers: 74.20.Fg, 71.45.Gm, 74.80.Dm, 78.30.-j}

\narrowtext

\section{INTRODUCTION}
Recently, the authors \cite{WGB3} studied the 
inelastic light-scattering intensity of a semi-infinite superconducting
superlattice with a bilayer basis. Motivated by the Cooper pair
tunneling model proposed by Chakravarty, Anderson, and coworkers
\cite{CSAS93} for high $T_c$ layered superconductors,
we discussed the in-phase and out-of-phase phase modes 
(corresponding to the phase fluctuations
of the two superconducting order parameters in a bilayer)
which arise in the presence of interlayer Cooper pair tunneling
\cite{WGB2}.  These modes  couple into density fluctuations and, as a result,
show up in the Raman inelastic light scattering. The 
intensity is weak because of screening associated 
with the Coulomb interaction, and in fact is below the threshold
of current Raman experiments. However, our results are of sufficient
interest that such experiments should be attempted.
In this paper, we extend our previous calculations \cite{WGB3}
and show how the Raman spectrum for $\omega < 2\Delta$
is significantly modified when we include the surface contribution.
It is also the first time surface plasmons (which occurs above $2\Delta$)
are included in the calculation of the 
Raman intensity of layered superconductors.

The present paper is based on the approach of
Jain and Allen \cite{JA}, who considered normal layered electron gas (LEG).
In their calculation for a semi-infinite superlattice,
both the bulk and surface contributions were included.
They found that there were two effects of the surface:
(1) Van Hove singularities at the upper and lower
limits of the bulk plasmon band were completely canceled out by
negative surface contributions. 
(2) Depending on the background dielectric constants,
surface plasmons \cite{GQ}
can appear, either above or below the bulk plasmon band.
We show that the analogous effects arise in a semi-infinite
{\em superconducting} superlattice, resulting in major modifications of
the bulk contribution to the
Raman scattering spectrum given in Ref.~\cite{WGB3}.
Apart from the out-of-phase phase mode contribution,
the Raman scattering intensity is found to be strongly dependent
on the value of momentum transfer $q_\parallel$ (parallel to the layers).
For simplicity, we only discuss the surface effects for a
superlattice with a single layer per unit cell. This is sufficient to 
understand the essential physics.

The isotropic inelastic light-scattering cross section is given by \cite{JA,SG}
\begin{equation}
{d\sigma \over d\omega d\Omega} \propto
|\hat{{\bf e}}_i \cdot \hat{{\bf e}}_f|^2
I({\bf q}, \omega),
\label{eq:Raman}
\end{equation}
where
\begin{eqnarray}
&\displaystyle I({\bf q},\omega)=\sum_
{\stackrel{\scriptstyle l,l^\prime}{i,j}}
{\rm Im}\chi_{ij}({\bf q}_\parallel,\omega,
l,l^\prime)e^{-(Z_{l,i}+Z_{l^\prime,j}) /\delta} &\nonumber\\
&\times e^{-2ik_\perp(Z_{l,i}-Z_{l^\prime,j})}.&
\label{eq:intensity}
\end{eqnarray}
Here $Z_{l,i}$ represents the position of the $i$-th
layer in the $l$-th unit cell.
The density response function $\chi_{ij}({\bf q}_\parallel,\omega,
l,l^\prime)$ in (\ref{eq:intensity})
has been evaluated in Ref~\cite{WGB3} and
represents the correlation between the charge density on layer $(l,i)$
and the charge density on layer $(l^\prime,j)$.  
The incident photon has momentum ${\bf k}_{i}$, energy $\omega_i$, and
polarization $\hat{{\bf e}}_i$ and the scattered photon is similarly
described by ${\bf k}_{f}$, $\omega_f$, and $\hat{{\bf e}}_f$.  
We assume that the
energy transfer $\omega\equiv \omega_i-\omega_f$ in (\ref{eq:intensity}) is
very small compared to the photon frequencies,  {\em i.e.}, $\omega_i\simeq
\omega_f$.  The momentum transfer parallel to the interface
is ${\bf k}_{\parallel,i}-{\bf k}_{\parallel,f}\equiv {\bf q}_\parallel$
and the momentum perpendicular to the interface is 
$k_{z,i}-k_{z,f}\equiv q_z$.  For small-angle scattering,
we have ${\rm Re}q_z\simeq 2k_\perp$ and ${\rm Im}q_z\simeq \delta^{-1}$
where $k_\perp$ is the momentum carried by the incident photon
and $\delta$ describes the damping of the photons in the medium.  
The result given in (\ref{eq:intensity}) shows
that as a result of the finite value of $\delta$,
the inelastic light-scattering cross section involves a {\em weighted} sum of
the correlation functions for electronic densities in the different layers.
We are interested in the interplay between the bulk and surface contributions
to the Raman spectra. For this purpose, we only keep the {\em isotropic}
matrix element for the Raman interaction given in
(\ref{eq:Raman}) \cite{Dev95}.  We consider superconductors with
{\it s}-wave layer pairing interaction but similar calculations
could be done for {\it d}-wave superconductors,
as discussed in Refs.~\cite{WGB2,Wu}.

\section{BULK IN-PHASE AND OUT-OF-PHASE PHASE MODES}
For later comparison, we first recall the results of
Ref.~\cite{WGB3} for the Raman spectra for a semi-infinite
superconducting {\em bilayer} superlattice, ignoring the surface contributions.
One finds that (\ref{eq:intensity}) reduces to
\begin{eqnarray}
I(\omega)&=&{1\over 1-e^{-2c/\delta}}{\rm Im}\Biggl\{
{E_+\over 2} \left(1+e^{-2d/\delta}+2e^{-d/\delta}\cos 2k_\perp d\right) 
\nonumber\\
&\times&\left[1+{2v_{\rm 2D}E_+\sinh q_{\parallel}c(u^2 e^{2c/\delta}-1)
\over F\sqrt{b^2-1}}\right]\nonumber\\
&+&{E_- \over 2D_{-}}
\left(1+e^{-2d/\delta}-2e^{-d/\delta}\cos 2k_\perp d\right)\Biggr\},
\label{eq:I.2}
\end{eqnarray}
where we have introduced the functions
\begin{eqnarray}
b&\equiv&\cosh q_\parallel c-2v_{\rm 2D}E_+ \sinh q_\parallel c \nonumber\\
u&\equiv&b+\sqrt{b^2-1}\nonumber\\
F&\equiv&u^2 e^{2c/\delta}-2u e^{c/\delta}\cos 2k_\perp c+1\nonumber\\
D_{-}&=& 1-v_{\rm 2D}(1-e^{-q_\parallel d})E_{-}.
\label{eq:abu}
\end{eqnarray}
The spacing of the bilayer is $d$, the unit cell length is $c$,
$v_{\rm 2D}\equiv 2\pi e^2 /q_\parallel\epsilon$ is
the 2D Coulomb interaction,
and $\epsilon$ is the superlattice background static dielectric constant.
In the long-wavelength limit ($q_\parallel\ll 2\Delta/v_F$), the functions 
$E_{\pm}({\bf q}_\parallel,\omega)$ are given by \cite{WGB2,WGB3}
\begin{equation}
E_{\pm}={1\over 4}N(\epsilon_F)J(\bar{\omega})
\left[{-R_\pm+{1\over 8}\bar{q}_\parallel^2 J(\bar{\omega}) \over
R_\pm+{1\over 4}(\bar{\omega}^2-{1\over 2}\bar{q}_\parallel^2)
J(\bar{\omega})}
\right],
\label{eq:E+-.final}
\end{equation}
where we have defined
\begin{eqnarray}
J(\bar{\omega})&=&\left\{\begin{array}{ll}
\displaystyle {2\over {\bar{\omega}\sqrt{1-\bar{\omega}^2}}}\arcsin\bar{\omega},
                 &~~\mbox{$\bar{\omega}<1$}\\
\displaystyle {2\over \bar{\omega}\sqrt{\bar{\omega}^2-1}}[\ln(\bar{\omega}-
           \sqrt{\bar{\omega}^2-1})+i{\pi \over 2}],
                 &~~\mbox{$\bar{\omega}>1$~,}
\end{array}\right.
\label{eq:I.s}
\end{eqnarray}
and
\begin{equation}
R_+=0~~~,~~~~~~ R_-={1\over gN(\epsilon_F)}{2x \over x^2-1}~~
;~~x\equiv {T_J\over g}.
\label{eq:def.R-1}
\end{equation}
Here $\bar{\omega}\equiv\omega/2\Delta$,
$\bar{q}_\parallel\equiv q_\parallel v_F/2\Delta$,
$N(\epsilon_F)\equiv m^*/\pi$ is the 2D electronic
density of states at the Fermi level
with $m^*$ being the effective electronic mass,
$g$ is the in-layer pairing interaction, and
$T_J$ is the interlayer Cooper pair tunneling strength.
Replacing $\omega\rightarrow  \omega+i\gamma$ is a simple way of
including finite energy-resolution.
On the rhs of (\ref{eq:I.2}), the first term ($\equiv I_I$)
gives the contribution from the in-phase phase fluctuations,
while the second term ($\equiv I_O$) is associated with the
out-of-phase phase fluctuations.

One finds that the in-phase first term
in (\ref{eq:I.2}) has {\em three} poles, given by
\begin{equation}
F(2k_\perp,q_\parallel,\omega)=0 \mbox{~~~~and~~~~}
b(q_\parallel,\omega)=\pm 1.
\label{eq:mode+}
\end{equation}
$F=0$ gives an in-phase plasmon mode 
which Raman scattering picks up
(in an approximate way \cite{JA}, this mode is similar
to the plasmon mode of an infinite superlattice, with
$q_z=2k_\perp$). The additional two (Van Hove) 
singularities given by the solutions of $b=\pm 1$ correspond 
to the upper ($+$) and lower ($-$) limits of the ``bulk plasmon'' band 
for an {\em infinite} superlattice \cite{JA}.
In contrast, the second term in (\ref{eq:I.2}) only has a 
single pole given by 
\begin{equation}
D_-(q_\parallel,\omega)=0,
\label{eq:D-=0}
\end{equation}
corresponding to out-of-phase phase mode discussed in detail in 
Ref.~\cite{WGB3}.
Because the unit cell summation is over many bilayers,
the Raman intensity is strongly enhanced in the superlattice case compared to
the isolated bilayer case (the latter is discussed in Ref.~\cite{Wu}). 
This is the origin of the
prefactor $[1-\exp(-2c/\delta)]^{-1}\approx \delta/2c$ in (\ref{eq:I.2}).
Using $c\equiv 12{\rm \AA}$ and $\delta\equiv 1000{\rm \AA}$, 
this prefactor is $\sim 40$.

We note in (\ref{eq:I.2}) that,
in the limit of $q_\parallel \rightarrow 0$, we have $R_+=0$ and hence
$E_+\sim \bar{q}_\parallel^2\rightarrow 0$.  This implies that the in-phase
phase modes given by $F=0$ and $b=\pm 1$ have 
less weight when $q_\parallel$ is small, an expected consequence of the
screening due to the Coulomb interaction.  In contrast,
even in the low-$q_\parallel$ limit, $E_-\sim R_-\sim x$ is still finite,
being proportional to the pair tunneling strength $T_J$.  This means that the
out-of-phase phase mode given by $D_-=0$ has a weight proportional to $x$
and is not too dependent on the value of $q_\parallel$
(in the range probed in Raman scattering experiments). In addition, one can 
see from (\ref{eq:I.2}) that the intensities of $I_I$ and $I_O$ are also
dependent on the factors ($1+e^{-2d/\delta}\pm 2e^{-d/\delta}\cos 2k_\perp
d$) which arise from the lattice summation  in (\ref{eq:intensity}).  Since
$d\ll \delta$ and $k_\perp d\ll 1$, we see that
the intensity for the out-of-phase phase modes ($-$ sign) is greatly
reduced compared to that of the in-phase phase modes ($+$ sign) as a result
of these factors.

In Fig.~\ref{fig:Raman.superlattice}, we 
plot the Raman light-scattering intensity based on (\ref{eq:I.2}).
In this and other figures, we use the
parameters:  bilayer spacing $d=3{\rm \AA}$, 
the unit cell size $c=12{\rm \AA}$, pairing strength
$gN(\epsilon_F)\equiv 0.25$, Fermi momentum $k_F=3.07\times 10^{7} {\rm
cm}^{-1}$ and hence a 
2D hole density $n=1.5\times10^{14} \/{\rm cm}^{-2}$, layer
effective mass $m^*=m$, background static dielectric constant $\epsilon=10$,
photon momentum in the $z$-direction $k_\perp =1.0\times 10^5 \/{\rm cm}^{-1}$,
the optical penetration depth $\delta\sim 1/k_\perp=1000{\rm \AA}$, the 
superconducting energy gap $\Delta=280{\rm cm}^{-1}$, 
and the finite-energy resolution $\gamma=0.05\Delta$.
The momentum transfer parallel to the layers is
$q_\parallel=5.0\times 10^{-3}\Delta/v_F=1.17\times 10^{3}\/{\rm cm}^{-1}$ 
in Fig.~\ref{fig:Raman.superlattice}.
The out-of-phase phase mode is well-defined, as expected.
While the Raman intensity from the out-of-phase contribution 
shown in Fig.~\ref{fig:Raman.superlattice} is not
too dependent on the value of $q_\parallel$,
roughly speaking, the intensity from the in-phase contribution
is proportional to $q_\parallel^2$.
As shown in Fig.~\ref{fig:Raman.superlattice} for the in-phase contributions, 
one has a small peak at $\omega=2\Delta$ corresponding to the 
pair-breaking gap in an {\it s}-wave superconductor (which is
identical to the upper limit of the bulk plasmon band, {\em i.e.}, the pole
given by $b=1$). 
In addition, ``hidden'' in the low-frequency broadened peak is 
the in-phase phase mode contribution given by the solution of
$F=0$, which overlaps on the Van Hove singularity
corresponding to the pole $b=-1$ at the lower limit of
the bulk superlattice plasmon band (see Fig.~\ref{fig:b=-1}).
This will become more transparent when we discuss the surface contribution.

In a normal metal superlattice \cite{JA}, 
the lower limit of the bulk plasmon band is
far away from the particle-hole continuum and, as a result, 
both the upper ($b=+1$ pole) and lower ($b=-1$ pole) limits of the bulk 
plasmon band are well-defined. We recall that the
bulk plasmon band refers to the plasmons labeled by $q_z$ in an {\em infinite}
superconducting superlattice. These give rise to the 
Van Hove singularities discussed in Ref.~\cite{JA}.  
In contrast, in a superconducting superlattice,
the particle-hole excitation  spectrum (which begins at the pair-breaking 
gap $2\Delta$) is strongly coupled into the superlattice 
bulk plasmon spectrum.  
As a consequence, the bulk plasmon band is split into two different 
regions above and below the pair-breaking gap ($2\Delta$).
For the plasmon band below $2\Delta$ (which we are most interested in),
one can find a well-defined line for $b=-1$ corresponding to the lower 
band limit of the bulk plasmon band (which is generally at low frequencies).
However, due to the strong coupling between the 
bulk plasmon and BCS particle-hole continuum, starting at $2\Delta$,
there is no well-defined solution for $b=+1$.
Nevertheless, the peak at $2\Delta$ in Fig.~\ref{fig:Raman.superlattice}
supports the argument that
$\omega=2\Delta$ can be considered as the effective upper limit of a 
superconducting bulk plasmon band \cite{CG,FD91}.

As Jain and Allen \cite{JA} have pointed out for a normal superlattice with
one layer per unit cell, the Van Hove singularity
associated with 
$b=+1$ corresponds to all the neighboring layers oscillating in-phase;
while the one associated with 
$b=-1$ corresponds to all the neighboring layers oscillating out-of-phase
with each other.  In contrast, the out-of-phase 
phase mode in Fig.~\ref{fig:Raman.superlattice}
is a  collective mode associated with
the ``internal Cooper pair dynamics'' exhibited by
a bilayer via the interlayer
Cooper pair tunneling between the two layers. The physics of
this out-of-phase phase mode is completely different from the 
out-of-phase $b=-1$ bulk plasmon in a superlattice. 

\section{SURFACE CONTRIBUTIONS AND SURFACE PLASMONS}
The result in (\ref{eq:I.2}) does not 
include the surface contribution. We now include it but 
only consider the case of a semi-infinite superlattice
of {\em single} layer per unit cell since this already describes the 
interplay between bulk and surface contributions.
In the case of a single layer per unit cell, we need only to replace
the usual 2D Lindhard function in the formulas given by
Jain and Allen \cite{JA} by the
appropriate density response function, [{\em i.e.}, 
$E_+$ in (\ref{eq:E+-.final})] for a {\em neutral} 2D superconductor \cite{WGB2}.
Using Eq.~(50) in Ref.~\cite{JA}, we find that
the resulting Raman intensity is given by
\begin{eqnarray}
I(\omega)&=&{1\over 1-e^{-2c/\delta}}{\rm Im}\Biggl\{ E_+\left[
\left(1+{v_{\rm 2D}E_+\sinh q_{\parallel}c(u^2 e^{2c/\delta}-1)
\over F\sqrt{b^2-1}}\right)\right.\nonumber\\
&+&\left.{v_{\rm 2D}E_+ (e^{2c/\delta}-1)
(u^2 A-2uB+C)\over 2Q(b^2-1)F}\right] \Biggr\},
\label{eq:I.2.surface}
\end{eqnarray}
where we have defined 
\begin{eqnarray}
A&\equiv&G\sinh^2 q_\parallel c+1+{\alpha\over 2}e^{2q_\parallel c},\nonumber\\
B&\equiv&H\sinh^2 q_\parallel c+\cosh q_\parallel c+{\alpha\over 2}
e^{q_\parallel c},\nonumber\\
C&\equiv&G\sinh^2 q_\parallel c+1+{\alpha\over 2},
\label{eq:ABC}
\end{eqnarray}
with
\begin{eqnarray}
G&\equiv&{1\over 2}[(b^2-1)^{-{1\over 2}}-1/\sinh q_\parallel c]
/\sinh q_\parallel c,\nonumber\\
H&\equiv&{1\over 2}[u^{-1}(b^2-1)^{-{1\over 2}}-e^{-q_\parallel c}/
\sinh q_\parallel c] /\sinh q_\parallel c,\nonumber\\
Q&\equiv&{1\over 2}[1-(b^2-1)^{-{1\over 2}}(1-b\cosh q_\parallel c)/
\sinh q_\parallel c]\nonumber\\
&&-{1\over 2}\alpha e^{q_\parallel c} (b^2-1)^{-{1\over 2}}
(\cosh q_\parallel c-b)/\sinh q_\parallel c,
\label{eq:GHQ}
\end{eqnarray}
where $u$, and $F$ are defined in (\ref{eq:abu}) and 
now $b=\cosh q_\parallel c-v_{\rm 2D}E_+ \sinh q_\parallel c$.
The parameter $\alpha\equiv (\epsilon-\epsilon_0)/(\epsilon+\epsilon_0)$
depends on the optical dielectric constants ($\epsilon$) inside and
($\epsilon_0$) outside the superlattice.
It plays a key role in determining the surface contributions as well as 
the appearance and energy of surface plasmons. The formula for $\alpha$
can be rewritten in the
useful form $\epsilon/\epsilon_0=(1+\alpha)/(1-\alpha)$.
We call attention to the similarity between the first term in 
(\ref{eq:I.2.surface}) and the first term in (\ref{eq:I.2}). 
In the rhs of (\ref{eq:I.2.surface}), the first term ($\equiv I_B$)
gives the {\em bulk} contribution, while the second term ($\equiv I_S$) 
is associated with the {\em surface} contribution.  The three 
poles mentioned earlier are exhibited by both contributions: 
$F=0$ corresponding to an in-phase plasmon which Raman
scattering picks up and $b=\pm 1$ corresponding to the upper and lower
limits of the bulk plasmon band. 
There is a new pole of the surface contribution $I_S$ 
given by 
\begin{equation}
Q(q_\parallel,\omega)=0,
\label{eq:Q=0}
\end{equation}
which can be shown to correspond to a surface plasmon. 
One has a nontrivial solution of $Q=0$ only 
when $\alpha\neq 0$ ({\em i.e.,} $\epsilon\neq \epsilon_0$), that is, the 
surface of the superconducting superlattice
must separate regions with different dielectric constants to
give rise to surface plasmons.

In Fig.~\ref{fig:b=-1}, we show the dispersion relation of 
surface plasmons in a superconducting superlattice, as given
by the solutions of $Q=0$, for various values of $\alpha$.
The shaded area between the line denoted by $b=-1$ to the line 
$\omega=2\Delta$ represents the bulk plasmon band of an infinite
superconducting superlattice. We find that the
surface plasmon appears only above the upper limit ({\em i.e.}, 
$\omega=2\Delta$) of this bulk plasmon band,
where BCS particle-hole damping of collective modes can occur. 
For positive values of $\alpha$ ($\epsilon>\epsilon_0$), a surface plasmon 
appears at very large energies $\omega\gg 2\Delta$,
in which case it is essentially identical to that in a normal superlattice.
We remark that for a high-$T_c$ material with a dielectric constant
$\epsilon\sim 10$ in a vacuum ($\epsilon_0=1$), one has $\alpha=0.82$.
For $\alpha$ increasingly negative ($\rightarrow -1.0$), 
which requires $\epsilon_0/\epsilon>1$, the surface plasmon energy slowly
{\em decreases} toward $2\Delta$ (see Fig.~\ref{fig:b=-1}).

In Fig.~\ref{fig:b=-1}, the dispersion relation denoted by $F=0$ 
represents a (bulk) plasmon, which is a pole of
the Raman scattering intensity in (\ref{eq:I.2.surface}).
One sees that $F=0$ mode is well-defined only when
$q_\parallel \alt  0.025\Delta/v_F$ (solid line). The critical value
of $q_\parallel$ ($0.025\Delta/v_F$) changes for different
choices of  $\delta$ and $k_\perp$. When 
$q_\parallel \agt 0.025\Delta/v_F$, we find no solution for $F=0$. 
The dashed line represents the minima of $F$ ({\em i.e.}, an over-damped or 
relaxational mode). This broad resonance is always
peaked at $\omega/2\Delta=0.8$, whatever the values chosen for
$\delta$ and $k_\perp$.

In Fig.~\ref{fig:cancel_1}, we show the 
net Raman intensity based on (\protect\ref{eq:I.2.surface}) 
for a semi-infinite superconducting single-layer superlattice,
showing the surface and the bulk components.
Comparing Fig.~\ref{fig:cancel_1} with Fig.~\ref{fig:Raman.superlattice},
one sees that the low-weight Van Hove singularity at $b=-1$ from
the bulk contribution is canceled by the negative surface
contribution. In contrast, the {\em in-phase} phase bulk plasmon mode 
($F=0$) shows up as a sharp peak in the
low-frequency region. As shown in Fig.~\ref{fig:cancel_1},
a peak associated with the $b=1$ Van Hove singularity
can still appear at $\omega=2\Delta$ since 
the bulk and surface contributions do not completely cancel each other.

Fig.~\ref{fig:cancel_2} is similar to 
Fig.~\ref{fig:cancel_1} using the same parameters, but at a much
higher momentum transfer $q_\parallel=0.1\Delta/v_F$.
The {\em cancellation} between the bulk and surface contributions is
clearly shown not only for the two boundaries of the bulk plasmon band
at $b=-1$ and $\omega=2\Delta$, but almost for the entire
region below $2\Delta$.  We also see that at 
$q_\parallel=0.1\Delta/v_F$, the broad relaxational mode 
[always peaked at $\omega\approx 0.8(2\Delta)$] corresponding to the
minimum of $F$ (see Fig.~\ref{fig:b=-1})
has only low weight in the Raman scattering spectrum.

To see how the interplay between the bulk and surface contribution
depends on $q_\parallel$, we plot in Fig.~\ref{fig:qinverse}
the {\em total} Raman intensities for various values of $q_\parallel$.
Comparing Fig.~\ref{fig:qinverse} with Fig.~\ref{fig:b=-1},
one finds that for $q_\parallel \alt  0.025\Delta/v_F$,
a well-defined bulk plasmon mode ($F=0$) gives rise to a sharp peak
even after the cancellation of the bulk and surface contributions.
A peak associated with $\omega=2\Delta$ still shows up.
As mentioned above, for $q_\parallel \agt 0.025\Delta/v_F$, 
the minimum of $F$ results in a very broad 
maximum at $\omega/2\Delta\approx 0.8$.  We also note that for
$q_\parallel \agt 0.025\Delta/v_F$,
while both the bulk and surface contributions are roughly proportional to
$q_\parallel^2$, the {\em net} spectrum after the mutual
cancellation of bulk and
surface contribution effectively {\em decreases} as $q_\parallel$ increases.

For $~c/\delta$, $~q_\parallel c$, and $~k_\perp c\ll 1$ appropriate for
layered superconductors, the total Raman intensity given in 
(\ref{eq:I.2.surface}) can be reduced after some calculation to
\begin{equation}
I(\omega)\simeq{\delta\over 2c}{\rm Im}\left( E_+\over F\right)
\left[\left({c\over \delta}\right)^2+(2k_\perp c)^2\right],
\label{eq:I.tot.approx}
\end{equation}
which is valid for $\bar{q_\parallel}\ll \bar{\omega} <1$ and for all
values of $\alpha$. We note that in (\ref{eq:I.tot.approx}),
the function $F$ given in (\ref{eq:abu}) is a very sensitive function of 
$c/\delta$ and $k_\perp c$ and therefore we cannot
approximate it by $(u-1)^2$ in the limit of $c/\delta$,  
$k_\perp c\rightarrow 0$. We can see directly from (\ref{eq:I.tot.approx}) 
that the poles of $b=\pm 1$ are removed as a consequence of
the cancellation between bulk and surface contributions. 
The only pole now is given by $F=0$.
One has a broad spectrum without any sharp peak
for  $q_\parallel \agt 0.025\Delta/v_F$.
In this region, one may verify that $F\propto E_+^2 \propto q_\parallel^4$.
As a result, the net Raman intensity
$I(\omega)$ decreases as $q_\parallel$ increases, roughly
proportional to $q_\parallel^{-2}$ (see Fig.~\ref{fig:qinverse}).
 
In addition to the surface contribution discussed above, there may be
a surface plasmon in the region $\omega>2\Delta$
(see Fig.~\ref{fig:b=-1}) but this only arises when we have
different dielectric constants inside and outside 
the superlattice ($\epsilon\neq\epsilon_0$).
In Fig.~\ref{fig:cancel_3}, we show the contribution of a
surface plasmon to the Raman intensity. The parameters used are as
in Fig.~\ref{fig:cancel_1} but at a much
higher momentum transfer $q_\parallel=0.2\Delta/v_F$ (which is probably
the upper limit for Raman scattering in high-$T_c$ superconductors).
In order to have a surface plasmon energy fairly close to 
$2\Delta$, the dielectric constants are taken to give $\alpha=-0.80$. 
The latter value requires $\epsilon_0/\epsilon=9$, {\it i.e.}, a layered 
superconductor with {\em much lower} dielectric constant
compared to the overlay material. 
As shown in Fig.~\ref{fig:cancel_3}, this gives a Raman spectrum with lots of
structure, with a broad surface plasmon
peak at an energy above $2\Delta$. This region is the pair-breaking region, 
where there is strong BCS particle-hole damping of collective modes.
The intensity of the surface plasmon is roughly 
proportional to $q_\parallel^2$. Once again, Fig.~\ref{fig:cancel_3}
shows the almost complete cancellation between 
the bulk and surface contributions in 
the region of $\omega\leq 2\Delta$.
This is because at this relatively large value of $q_\parallel$,
there is no well-defined solution of $F=0$ (see Fig.~\ref{fig:b=-1}).

One might worry that the bilayer
out-of-phase phase mode shown in Fig.~\ref{fig:Raman.superlattice}
might also be strongly modified
due to the surface contributions at higher values of $q_\parallel$.
However, while the intensity (and the dispersion relation) of the
out-of-phase phase mode is very sensitive to the pair tunneling strength
$T_J$, it is not strongly dependent on the value of $q_\parallel$.
It is associated with out-of-phase oscillation of order parameters
in a single bilayer, with 
no {\em net} charge fluctuation \cite{WGB3}. Therefore, we have no
reason to expect any strong modification of the out-of-phase phase mode
at larger values of $q_\parallel$.

\section{CONCLUSIONS}
We have shown that the surface contribution plays a major role in
determining the final Raman-spectrum line shape
from semi-infinite superconducting superlattices. 
We find that, as in the case of
normal superlattices discussed by Jain and Allen \cite{JA},
the proper inclusion of the surface contribution cancels ``spurious''
bulk contributions associated with Van Hove singularities 
(see Ref.~\cite{WGB3}) of the upper and lower limits of
the bulk plasmon band of an infinite superconducting superlattice. 
We have also found that the bulk plasmon mode in the
region $\omega<2\Delta$ ceases to be well-defined when 
$q_\parallel$ reaches a critical value ($\sim 0.025\Delta/v_F$ for the 
parameters we have used).
In addition, the surface plasmon 
usually contributes to the Raman
intensity in the region well {\em above} $2\Delta$ and
only approaches $2\Delta$ if the superconductor is
overlayed by a transparent material with a dielectric
constant $\epsilon_0$ much larger than that of the superlattice $\epsilon$.
The surface-plasmon mode intensity increases as $q_\parallel^2$ and
thus one wants $q_\parallel$ as large as possible if one wants to study it
(see Fig.~\ref{fig:cancel_3}).
As mentioned above, for relatively large values of $q_\parallel$,
the (negative) surface contribution tends to completely cancel out the bulk 
contribution in the entire frequency region {\em below} $2\Delta$.
Because the out-of-phase phase mode discussed in Ref.~\cite{WGB3}
is not expected to be strongly affected by the surface contribution,
at relatively large values of $q_\parallel$, this mode 
(see Fig.~\ref{fig:Raman.superlattice}) 
should be the only structure remaining in the Raman spectrum
below $2\Delta$.

The absolute intensity of the Raman spectra
we discuss in this paper is somewhat below current experimental sensitivity.
However we hope that the interesting predictions we make
concerning the out-of-phase phase modes as 
well as surface plasmons will encourage future experimental efforts.

\acknowledgements
We thank Bryan Statt for first asking about the role of surface plasmons.
This work was supported by a research grant from NSERC of Canada.



\begin{figure}
\caption{The bulk Raman intensity \protect\cite{statement1}
given by (\protect\ref{eq:I.2}) 
for a semi-infinite superconducting bilayer superlattice with interlayer 
Cooper-pair tunneling of amplitude $T_J$.
We use $T_J=0.03 g$ ($x=0.03$) and $q_\parallel=5.0\times 10^{-3}\Delta/v_F$.
The surface contribution is not included.} 
\label{fig:Raman.superlattice}
\end{figure}

\begin{figure}
\caption{The dispersion relation of surface plasmons in a semi-infinite
superconducting superlattice (solutions of $Q=0$)
for $\alpha=-0.8$ and $-0.9$.
The shaded area between $b=-1$ and $\omega=2\Delta$ lines
denotes the bulk plasmon band for an infinite superconducting superlattice.
The $F=0$ line represents a sharp ``bulk plasmon'' picked up 
by Raman scattering. For $q_\parallel \protect\agt 0.025\Delta/v_F$ 
(dashed line),
there is only an over-damped resonance (see Fig.~\protect\ref{fig:qinverse}).}
\label{fig:b=-1}
\end{figure}

\begin{figure}
\caption{Raman intensity given by (\protect\ref{eq:I.2.surface}) 
from both bulk (dotted line) and surface (dashed line) 
contributions for a semi-infinite superconducting superlattice.
The same parameters are used as in Fig.~\protect\ref{fig:Raman.superlattice},
with $\alpha=0.82$ (corresponding to $\epsilon_0=1$ and $\epsilon=10$).
Apart from the absence of the out-of-phase phase mode
(only expected when the unit cell has two layers),
the bulk contribution is very similar to that given in 
Fig.~\protect\ref{fig:Raman.superlattice}.}
\label{fig:cancel_1}
\end{figure}

\begin{figure}
\caption{Raman intensities as in Fig.~\protect\ref{fig:cancel_1}
except for a much larger value $q_\parallel=0.1\Delta/v_F$.
This plot clearly shows the almost complete mutual cancellation of
the bulk and surface contributions at such large values of $q_\parallel$.}
\label{fig:cancel_2}
\end{figure}

\begin{figure}
\caption{The total Raman intensity in the region $\omega\leq 2\Delta$
(including {\em both} bulk and surface
contributions) for various values of the momentum transfer $q_\parallel$.
The peak is associated with the zero or minimum of $F$ in 
(\protect\ref{eq:I.tot.approx}).} 
\label{fig:qinverse}
\end{figure}

\begin{figure}
\caption{The Raman intensity as in Fig.~\protect\ref{fig:cancel_1},
with $q_\parallel=0.2\Delta/v_F$ and $\alpha=-0.8$.~~~~~~~~~~~~~~~~~~}
\label{fig:cancel_3}
\end{figure}

\end{document}